\documentclass[journal]{IEEEtran}

\ifCLASSINFOpdf

\else

\fi

\usepackage{cite}
\usepackage{graphicx}
\usepackage{subfigure}
\usepackage{amsmath}
\usepackage{amssymb}
\usepackage{url}
\usepackage{algorithm}
\usepackage{algorithmic}
\usepackage{multirow}
\begin{document}

\title{A Survey of Deep Reinforcement Learning in Video Games}

\author{Kun~Shao,
        Zhentao~Tang,
        Yuanheng~Zhu,~\IEEEmembership{Member,~IEEE},
        Nannan~Li,
        and Dongbin~Zhao,~\IEEEmembership{Fellow,~IEEE}
        % <-this % stops a space
\thanks{K. Shao, Z. Tang, Y. Zhu, N. Li, and D. Zhao are with the State Key Laboratory of Management and Control for Complex Systems, Institute of Automation, Chinese Academy of Sciences. Beijing 100190, China. They are also with the University of Chinese Academy of Sciences, Beijing, China (e-mail: shaokun2014@ia.ac.cn; tangzhentao2016@ia.ac.cn; yuanheng.zhu@ia.ac.cn; linannan2017@ia.ac.cn, dongbin.zhao@ia.ac.cn).}
\thanks{This work is supported by National Natural Science Foundation of China (NSFC) under Grants No.61573353, No.61603382, No.6180337, and No.61533017.}
}

\maketitle

\begin{abstract}

%Deep reinforcement learning (DRL) has made rapid development in the last few years, especially in video games.
Deep reinforcement learning (DRL)  has made great achievements since proposed.
Generally, DRL agents receive high-dimensional inputs at each step, and make actions according to deep-neural-network-based policies.
This learning mechanism updates the policy to maximize the return with an end-to-end method.
In this paper, we survey the progress of DRL methods, including value-based, policy gradient, and model-based algorithms, and compare their main techniques and properties. Besides, DRL plays an important role in game artificial intelligence (AI). We also take a review of the achievements of DRL in various video games, including classical Arcade games, first-person perspective games and multi-agent real-time strategy games, from 2D to 3D, and from single-agent to multi-agent.
A large number of video game AIs with DRL have achieved super-human performance, while there are still some challenges in this domain.
Therefore, we also discuss some key points when applying DRL methods to this field, including exploration-exploitation, sample efficiency, generalization and transfer, multi-agent learning, imperfect information, and delayed spare rewards, as well as some research directions.

\end{abstract}

\begin{IEEEkeywords}
reinforcement learning, deep learning, deep reinforcement learning, game AI, video games.
\end{IEEEkeywords}

\IEEEpeerreviewmaketitle

\section{Introduction}

\IEEEPARstart{A}{rtificial} intelligence (AI) in video games is a long-standing research area.
It studies how to use AI technologies to achieve human-level performance when playing games. More generally, it studies the complex interactions between agents and game environments.
Various games provide interesting and complex problems for agents to solve, making video games perfect environments for AI research.
These virtual environments are safe and controllable.
In addition, these game environments provide infinite supply of useful data for machine learning algorithms, and they are much faster than real-time.
These characteristics make games the unique and favorite domain for AI research.
On the other side, AI has been helping games to become better in the way we play, understand and design them \cite{Yannakakis2018AI}.

Broadly speaking, game AI involves the perception and the decision-making in game environments.
With these components, there are some crucial challenges and proposed solutions.
The first challenge is that the state space of the game is very large, especially in strategic games.
With the rise of representation learning, the whole system has successfully modeled large-scale state space with deep neural networks.
The second challenge is that learning proper policies to make decisions in dynamic unknown environment is difficult. For this problem, data-driven methods, such as supervised learning and reinforcement learning (RL), are feasible solutions.
The third challenge is that the vast majority of game AI is developed in a specified virtual environment.
How to transfer the AI's ability among different games is a core challenge.
A more general learning system is also necessary.

\begin{figure*}[!t]
\centering
\includegraphics[width=5.8 in]{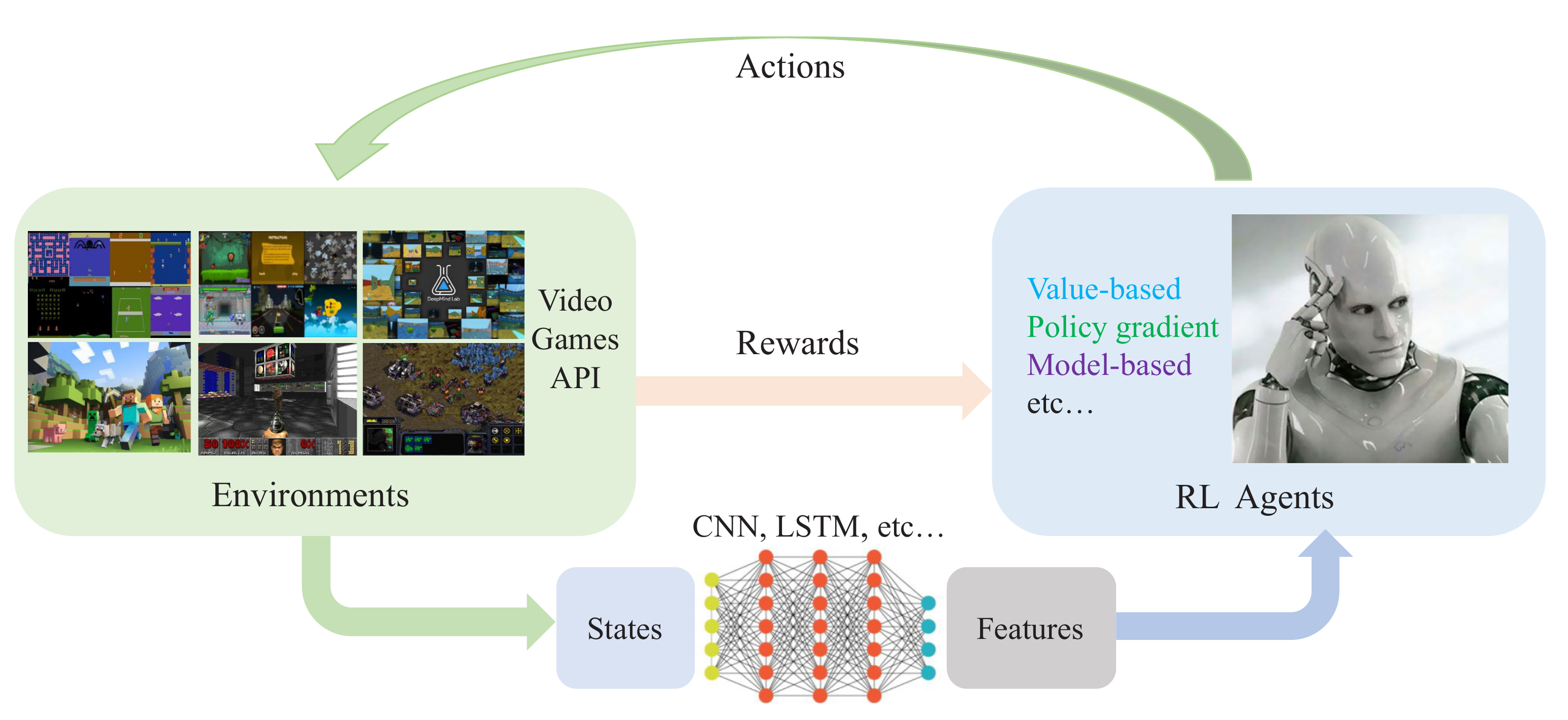}
\caption{The framework diagram of the typical DRL for video games. The deep learning model takes input from video games API, and extract meaningful features automatically. DRL agents produces actions based on these features, and make the environments transfer to next state.}
\label{fig_sim}
\end{figure*}

For a long time, solving these challenges with reinforcement learning is widely used in game AI.
And in the last few years, deep learning (DL) has achieved remarkable performance in computer vision and natural language processing \cite{Lecun2015Deep}.
The combination, deep reinforcement learning (DRL), teaches agents to make decisions in high-dimensional state space in an end-to-end framework, and dramatically improves the generalization and scalability of traditional RL algorithms.
Especially, DRL has made great progress in video games, including Atari, ViZDoom, StarCraft, Dota2, and so on.
There are some related works to introduce these achievements in this field.
Zhao et al. \cite{Zhao2016Review} and Tang et al. \cite{tang2018Recent} survey the development of DRL research, and focus on AlphaGo and AlphaGo Zero.
Justesen et al. \cite{Justesen2017DeepLF} reviews  DL-based methods in video game play, including supervised learning, unsupervised learning, reinforcement learning, evolutionary approaches, and some hybrid approaches.
Arulkumaran et al. \cite{Arulkumaran2017DeepRL} make a brief introduction of DRL, covering central algorithms and presenting a range of visual RL domains.
Li \cite{Li2017DeepRL} gives an overview of recent achievements of DRL, and discusses core elements, important mechanisms, and various applications.
In this paper, we focus on DRL-based game AI, from 2D to 3D, and from single-agent to multi-agent.
The main contributions include the comprehensive and detailed comparisons of various DRL methods, their techniques, properties, and the impressive and diverse performances in these given video games.

The organization of the remaining paper is arranged as follows.
In Section II, we introduce the background of DL and RL.
In Section III, we focus on recent DRL methods, including value-based, policy gradient, and model-based DRL methods.
After that, we make a brief introduction of research platforms and competitions, and present performances of DRL methods in classical single-agent Arcade games, first-person perspective games, and multi-agent real-time strategy games.
In Section V, we discuss some key points and research directions in this field.
In the end, we draw a conclusion of this survey.

\section{Background}

Generally speaking, training an agent to make decisions with high-dimensional inputs is difficult. With the development of deep learning, researchers take deep neural networks as function approximations, and use plenty of samples to optimize policies successfully. The framework diagram of typical DRL for video games is depicted in Fig. 1.

\subsection{Deep learning}
Deep learning comes from artificial neural networks, and is used to learn data representation. It is inspired by the theory of brain development, and can be learned in supervised learning, unsupervised learning and semi-supervised learning. Although the term deep learning is introduced in 1986 \cite{dechter1986learning}, deep learning has a winter time because of lacking data and incapable computation hardware. However, with more and more large-scale datasets being released, and capable hardware being available, a big revolution happens in DL \cite{schmidhuber2015deep}.

Convolutional neural network (CNN) \cite{Krizhevsky2012ImageNet} is a class of deep neural networks, which is widely applied to computer vision.
CNN is inspired by biological processes, and is shift invariant based on shared-weights architecture.
Recurrent Neural Network (RNN) is another kind of deep nerial network, especially for natural language processing.
As a special kind of RNN, Long Short Term Memory (LSTM) \cite{hochreiter1997long} is capable of learning long-term dependencies.
%At present, LSTM is widely used and work tremendously well on a large variety of problems.
Deep learning architectures have been applied into many fields, and have achieved significant successes, such as speech recognition, image classification and segmentation, semantic comprehension, and machine translation \cite{Lecun2015Deep}.
DL-based methods with efficient parallel distributed computing resources can break the limit of traditional machine learning methods. This method inspires scientists and researchers to achieve more and more state-of-the-art performance in respective fields.

\begin{figure*}[!t]
\centering
\includegraphics[width=7.2 in]{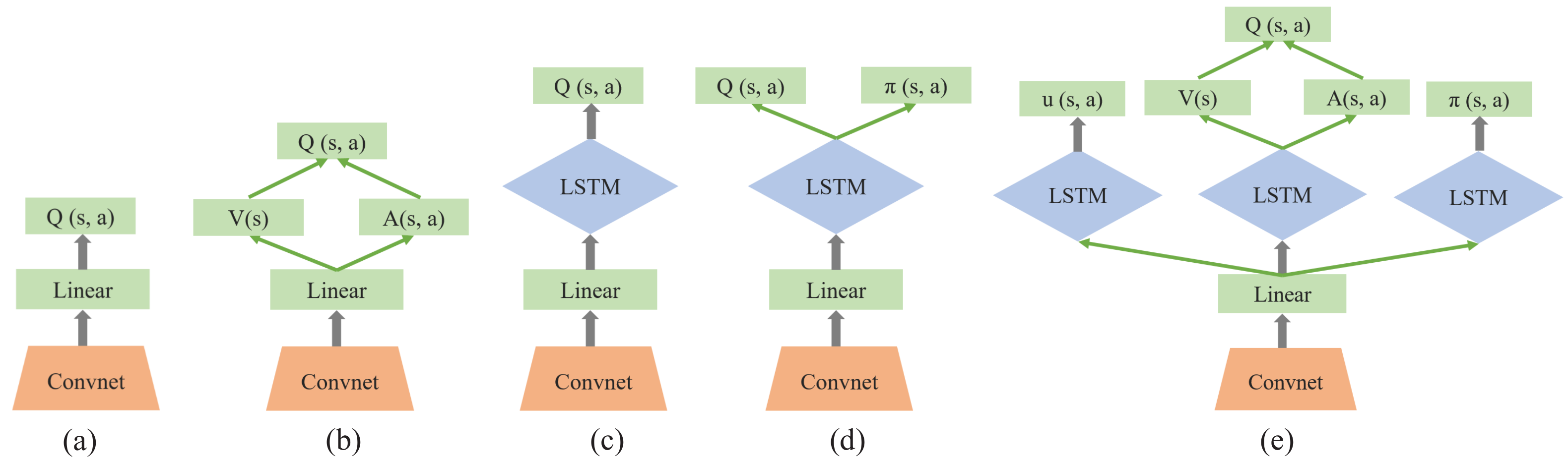}
\caption{The network architectures of typical DRL methods, with increased complexity and performance. (a): DQN network; (b)Dueling DQN network; (c): DRQN network; (d): Actor-critic network; (e): Reactor network.}
\label{fig_sim}
\end{figure*}

\subsection{Reinforcement learning}
Reinforcement learning is a kind of machine learning methods where agents learn the optimal policy by trial and error \cite{Sutton1998Reinforcement}. By interacting with the environment, RL can be successfully applied to sequential decision-making tasks.
Considering a discounted episodic Markov decision process (MDP) $(S,A,\gamma,P,r)$, the agent chooses an action $a_t$ according to the policy $\pi(a_t|s_t)$ at state $s_t$. The environment receives the action, produces a reward $r_{t+1}$ and transfers to the next state $s_{t+1}$ according to the transition probability $P (s_{t+1}|s_t, a_t)$. This transition probability is unknown in RL domain. The process continues until the agent reaches a terminal state or a maximum time step.
The objective is to maximize the expected discounted cumulative rewards
\begin{equation}
\mathbb{E}_\pi[R_t] = \mathbb{E}_\pi[\sum_{i=0}^\infty \gamma^i r_{t+i}] ,
\end{equation}
where $\gamma \in(0,1]$ is the discount factor.

Reinforcement learning can be devided into off-policy and on-policy methods.
Off-policy RL algorithms mean that the behavior policy used for selecting actions is different from the learning policy.
On the contrary, behavior policy is the same with the learning policy in on-policy RL algorithms.
Besides, reinforcement learning can also be devided into value-based and policy-based methods.
In value-based RL, agents update the value function to learn suitable policy, while policy-based RL agents learn the policy directly.

Q-learning is a typical off-policy value-based method. The update rule of Q-learning is
\begin{subequations}
\begin{gather}
\delta_t = r_{t+1} + \gamma \arg\max_a Q(s_{t+1},a) - Q(s_t,a_t) , \\
Q(s_t,a_t) \leftarrow Q(s_t,a_t) + \alpha\delta_t .
\end{gather}
\end{subequations}
$\delta_t$ is the temporal difference (TD) error, and $\alpha$ is the learning rate.

Policy gradient \cite{Williams1992SimpleSG} parameterizes the policy and updates parameters $\theta$.
In its general form, the objective function of policy gradient is defined as
\begin{equation}
J(\theta) = \mathbb{E}_\pi[\sum_{t=0}^\infty \log \pi_{\theta}(a_t|s_t) R].
\end{equation}
$R$ is the total accumulated return.

Actor-critic \cite{Sutton1998Reinforcement} reinforcement learning improves the policy gradient with an value-based critic
\begin{equation}
J(\theta) = \mathbb{E}_\pi[\sum_{t=0}^\infty \Psi_t \log \pi_{\theta}(a_t|s_t)].
\end{equation}
$\Psi_t$ is the critic, which can be the state-action value function $Q^\pi(s_t, a_t)$, the advantage function $A^\pi(s_t, a_t) = Q^\pi(s_t, a_t) - V^\pi(s_t)$ or the TD error $r_t + V^\pi(s_{t+1})-V^\pi(s_t)$.

\section{Deep Reinforcement Learning}

DRL makes a combination of DL and RL, achieving rapid developments since proposed. This section will introduce various DRL methods, including value-based methods, policy gradient methods, and model-based methods.

\subsection{Value-based DRL methods}

%\subsubsection{Value expectationl}

Deep Q-network (DQN) \cite{Mnih2015Human} is the most famous DRL model which learns policies directly from high-dimensional inputs. It receives raw pixels, and outputs a value function to estimate future rewards, as shown in Fig. 2(a). DQN uses the experience replay method to break the sample correlation, and stabilizes the learning process with a target Q-network.
The loss function at iteration $i$ is
\begin{eqnarray}
\begin{aligned}
L_i(\theta_i) = E_{(s,a,r,s')\sim U(D)}[(y_i^{DQN} - Q(s,a;{\theta _i}))^2],
\end{aligned}
\end{eqnarray}
with
\begin{equation}
y_i^{DQN} = r + \gamma \mathop {\max }\limits_{a'}Q(s',a';{\theta_i{^{-}}}).
\end{equation}

DQN bridges the gap between high-dimensional visual inputs and actions.
%, resulting in the first agent that is capable of learning to play a series of challenging video games.
After that, researchers have improved DQN in different aspects.
Double DQN \cite{Hasselt2016DeepRL} introduces double Q-learning to reduce observed overestimations, and it leads to much better performance.
%In previous DQN, experience transitions are uniformly sampled from a replay memory, regardless of their significance.
Prioritized experience replay \cite{Schaul2015PrioritizedER} helps prioritize experience to replay important transitions more frequently.
The sample probability of transition $i$ as $P(i)= \frac{p_i^{\alpha}}{\sum_k p_k^{\alpha}}$, where $p_i$ is the priority of transition $i$.
%Prioritized experience replay approach is related to the importance sampling techniques, and focuses learning on the most valuable experiences to learn more efficiently.
Dueling DQN \cite{Wang2016DuelingNA} uses the dueling neural network architecture for model-free DRL.
It includes two separate estimators: one for state value function $V(s; \theta, \beta)$ and the other for advantage function $A(s, a; \theta, \alpha)$, as shown in Fig. 2(b).
\begin{equation}
Q(s,a:\theta, \alpha, \beta)= V(s; \theta, \beta) + A(s, a; \theta, \alpha) .
\end{equation}

Pop-Art \cite{Hasselt2016LearningVA} is proposed to adapt to different and non-stationary target magnitudes, which successfully replaces the clipping of rewards as done in DQN to handle various magnitudes of targets.
%\cite{bellemare2016increasing} introduces new optimality-preserving operators on Q-functions.
%%Extending the idea of a locally consistent operator,
%The authors derive sufficient conditions for an operator to preserve optimality, leading to a family of operators which includes the consistent Bellman operator.
%They provide a proof of optimality for advantage learning algorithm and derive other gap-increasing operators with interesting properties.
Fast reward propagation \cite{He2016LearningTP} is a novel training algorithm for reinforcement learning, which combines the strength of DQN, and exploits longer state-transitions in experience replays by tightening the optimization via constraints.
This novel technique makes DRL more practical by drastically reducing training time.
Gorila \cite{Nair2015MassivelyPM} is the first massively distributed architecture for DRL.
This architecture uses four main components: parallel actors; parallel learners; a distributed neural network to represent the value function or behavior policy; and a distributed store of experience.
%DQN uses a convolutional neural network to encoder the visual inputs, followed by fully connected layers with a single output for each action.
%This architecture helps to generalize learning across actions, and leads to better policy evaluation with many similar-valued actions.
To address the limited memory and imperfect game information at each decision point, Deep Recurrent Q-Network (DRQN) \cite{Hausknecht2015DeepRQ} replaces the first fully-connected layer with a recurrent neural network in DQN, as shown in Fig. 2(c).
%The resulting Deep Recurrent Q-Network (DRQN) successfully integrates information through time.
%\cite{Sorokin2015DeepAR} presents an extension of DQN by attention mechanisms, termed as Deep Attention Recurrent Q-Network (DARQN).
%DARQN highlights the regions of the game screen that the agent is focusing on when making decisions during the training process.

Generally, DQN learns rich domain representations and approximates the value function with deep neural networks,
while batch RL algorithms with linear representations are more stable and require less hyperparameter tuning.
The Least Squares DQN (LS-DQN) \cite{Levine2017ShallowUF} combines DQN's rich feature representations with the stability of a linear least squares method.
In order to reduce approximation error variance in DQN‘s target values, averaged-DQN \cite{Anschel2017AveragedDQNVR} averages previous Q-values estimates, leading to a more stable training and improved performance.
Deep Q-learning from Demonstrations (DQfD) \cite{hester2017deep} combines DQN with human demonstrations, which improves the sample efficiency greatly.
DQV \cite{sabatelli2018deep} uses TD learning to train a Value neural network, and uses this network to train a second Quality-value network to estimate state-action values. DQV learns significantly faster and better than double-DQN.
Researchers have proposed several improvements to DQN. However, it is unclear which of these are complementary and how much can be combined. Rainbow \cite{Hessel2018RainbowCI} combines with main extensions to DQN, and gives each component's  contribution to overall performance.
RUDDER \cite{ArjonaMedina2018RUDDERRD} is a novel reinforcement learning approach for finite MDPs with delayed rewards, which is also a return decomposition method,
RUDDER is exponentially faster on tasks with different lengths of reward delays.
Ape-X DQfD\cite{Pohlen2018ObserveAL} uses a new transformed Bellman operator to process rewards of varying densities and scales, and applies human demonstrations to ease the exploration problem to guide agents towards rewarding states. Additional, it proposes an auxiliary temporal consistency loss to train stably extending the effective planning horizon by an order of magnitude.
Soft DQN \cite{Schulman2017EquivalenceBP} is an entropy-regularized versions of Q-learning, with better robustness and generalization
.

%\subsubsection{Value distributional}

Distributional DRL learns the value distribution, in contrast to common RL that models the expectation of return, or value.
C51 \cite{Bellemare2017A} focuses on the distribution of value, and designs distributional DQN algorithm to learn approximate value distributions.
QR-DQN \cite{Dabney2018DistributionalRL} methods close a number of gaps between theoretical and algorithmic results.
Distributional reinforcement learning with Quantile regression in which the distribution over returns is modeled explicitly instead of only estimating the mean.
Implicit Quantile Networks (IQN) \cite{Dabney2018ImplicitQN} is a flexible, applicable, and state-of-the-art distributional DQN. IQN approximates the full Quantile function for the return distribution with Quantile regression, and provides a fully integrated distributional RL agent without prior assumptions on the parameterization of the return distribution. Furthermore, IQN allows to expand the class of control policies to a wide range of risk-sensitive policies connected to distortion risk measures.

\subsection{Policy gradient DRL methods}

Policy gradient DRL optimizes the parameterized policy directly.
Actor-critic architecture computes the policy gradient using a value-based critic function to estimate expected future reward, as shown in Fig. 2(d).
Asynchronous DRL is an efficient framework for DRL that uses asynchronous gradient descent to optimize the policy\cite{Mnih2016AsynchronousMF}.
Asynchronous advantage actor-critic (A3C) trains several agents on multiple environments, showing a stabilizing effect on training. The objective function of the actor is demonstrated as
\begin{equation}
J(\theta) = \mathbb{E}_\pi[\sum_{t=0}^\infty A_{\theta, \theta_v}(s_t, a_t)\log\pi_{\theta}(a_t|s_t) + \beta H_{\theta}(\pi(s_t))],
\end{equation}
where $H_{\theta}(\pi(s_t))$ is an entropy term used to encourage exploration.

GA3C \cite{Babaeizadeh2016ReinforcementLT} is a hybrid CPU/GPU version of A3C, which achieves a significant speed up compared to the original CPU implementation.
UNsupervised REinforcement and Auxiliary Learning (UNREAL) \cite{Jaderberg2017ReinforcementLW} learns separate policies for maximizing many other pseudo-reward functions simultaneously, including value function replay, reward prediction, and pixel control.
This agent drastically improves both data efficiency and robustness to hyperparameter settings.
PAAC \cite{Clemente2017EfficientPM} is a novel framework for efficient parallelization of DRL, where multiple actors learn the policy on a single machine.
%The framework is algorithm agnostic and can be applied to on-policy, off-policy, value-based and policy gradient algorithms.
Policy gradient methods are efficient techniques for policies improvement, while they are usually on-policy and unable to take advantage of off-policy data. The new method is referred as PGQ\cite{ODonoghue2016PGQCP}, which combines policy gradient with Q-learning.
PGQ establishes an equivalency between regularized policy gradient techniques and advantage function learning algorithms.
Retrace($\lambda$) \cite{Munos2016SafeAE} takes the best of the importance sampling, off-policy Q($\lambda$), and tree-backup($\lambda$), resulting in low variance, safety, and efficiency. It makes a combination of dueling DRQN architecture and actor-critic architecture, as shown in Fig. 2(e).
Reactor \cite{Gruslys2017TheRA} is a sample-efficient and numerical efficient reinforcement learning agent based on a multi-step return off-policy
actor-critic architecture. The network outputs a target policy, an action-value Q-function, and an estimated behavioral policy. The critic is trained with the off-policy multi-step Retrace method and the actor is trained by a $\beta$-leave-one-out policy gradient.
Importance-Weighted Actor Learner Architecture (IMPALA) \cite{Espeholt2018IMPALASD} is a new distributed DRL, which can scale to thousands of machine. IMPALA uses a single reinforcement learning agent with a single set of parameters to solve a mass of tasks.
This method achieves stable learning by combining decoupled acting and learning with a novel V-trace off-policy
correction method, which is critical for achieving learning stability.

\subsubsection{\textbf{Trust region method}}
Trust Region Policy Optimization (TRPO) \cite{Schulman2015TrustRP} is proposed for optimizing control policies, with guaranteed monotonic improvement.
TRPO computes an ascent direction to improve on policy gradient, which can ensure a small change in the policy distribution.
The constrained optimization problem of TRPO in each epoch is
\begin{subequations}
\begin{gather}
maximize_{\theta} \ \ \  E_{s\sim\rho_{\theta'}, a\sim\pi_{\theta'}}[\frac{\pi_{\theta}(a|s)}{\pi_{\theta'}(a|s)}A_{\theta'}(s,a)], \\
s.t. \ \ \ \ E_{s \sim \rho_{\theta'}}[D_{KL}(\pi_{\theta'}(\cdot|s))] \leq \delta_{KL}.
\end{gather}
\end{subequations}
This algorithm is effective for optimizing large nonlinear policies.
Proximal policy optimization (PPO) \cite{Schulman2017ProximalPO} samples data by interaction with the environment, and optimizes the objective function with stochastic gradient ascent
\begin{subequations}
\begin{gather}
r_t(\theta) = \frac{\pi_\theta(a_t|s_t)}{\pi_{\theta_{old}}(a_t|s_t)} ,\\
L(\theta) = \hat{\mathbb{E}}_t[min(r_t(\theta)\hat{A}_t, clip(r_t(\theta), 1-\epsilon, 1+\epsilon)\hat{A}_t] .
\end{gather}
\end{subequations}
$r_t(\theta)$ denotes the probability ratio.
This objective function clips the probability ratio to modify the surrogate objective.
PPO has some benefits over TRPO, and is much simpler to implement, with better sample complexity.
Actor-critic with experience replay (ACER) \cite{Wang2016SampleEA} introduces several innovations, including stochastic dueling network, truncated importance sampling, and a new trust region method, which is stable and sample efficient.
Actor-critic using Kronecker-Factored Trust Region (ACKTR) \cite{Wu2017ScalableTM} bases on natural policy gradient, and uses Kronecker-factored approximate curvature (K-FAC) with trust region to optimize the actor and the critic. ACKTR is sample efficient compared with other actor-critic methods.

\subsubsection{\textbf{Deterministic policy}}
Apart from stochastic policy, deep deterministic policy gradient (DDPG) \cite{Lillicrap2015ContinuousCW} is a kind of deterministic policy gradient method which adapts the success of DQN to continuous control.
The update rule of DDPG is
\begin{equation}
Q(s_t,a_t) = r(s_t,a_t) + \gamma Q(s_{t+1},\pi_{\theta} (s_{t+1})).
\end{equation}
DDPG is an actor-critic, off-policy algorithm, and is able to learn reasonable policies on various tasks.
Distributed Distributional DDPG (D4PG) \cite{BarthMaron2018DistributedDD} is a distributional update to DDPG, combined with the use of multiple distributed workers all writing into the same replay table. This method has a much better performance on a number of difficult continuous control problems.

\subsubsection{\textbf{Entropy-regularized policy gradient}}
Soft Actor Critic (SAC) is an off-policy policy gradient method, which establishes a bridge between DDPG and stochastic policy optimization. SAC incorporates the clipped double-Q trick, and the objective function of maximum entropy DRL is
\begin{equation}
J(\pi) = \sum_{t=0}^T \mathbb{E}_{(s_t, a_t) \sim \rho_{\pi}} [ r(s_t, a_t) + \alpha H(\pi(.|s_t))],
\end{equation}

SAC uses an entropy regularization in its objective function. It trains the policy to maximize a trade-off between entropy and expected return. The entropy is a measure of randomness in the policy. This mechanism is similar to the  trade-off between exploration and exploitation. Increasing entropy can encourage more exploration, and accelerate learning process. Moreover, it can also prevent the learning policy from converging to a poor local optimum.

\subsection{Model-based DRL methods}

Combining model-free reinforcement learning with on-line planning is a promising approach to solve the sample efficiency problem.
TreeQN \cite{Farquhar2017TreeQNAA} is proposed to address these challenges. It is a differentiable, recursive, tree-structured model that serves as a drop-in replacement for any value function network in DRL with discrete actions.
TreeQN dynamically constructs a tree by recursively applying a transition model in a learned abstract state space and then aggregating predicted rewards and state-values using a tree backup to estimate Q-values.
ATreeC is an actor-critic variant that augments TreeQN with a softmax layer to form a stochastic policy network. Both approaches are trained end-to-end, such that the learned model is optimized for its actual use in the planner. TreeQN and ATreeC outperform n-step DQN and value prediction networks on multiple Atari games.
Vezhnevets et al. \cite{Vezhnevets2016StrategicAW} presents STRategic Attentive Writer (STRAW) neural network architecture to build implicit plans. STRAW purely interacts with an environment, and is an end-to-end method.
%The network builds an internal plan, which is continuously updated upon observation of the next input from the environment. Combining these properties,
STRAW model can learn temporally abstracted high-level macro-actions, which enables both economic computation and structured exploration. STRAW employs temporally extended planning strategies and achieves strong improvements on  Atari games.
The world model \cite{ha2018recurrent} uses an unsupervised manner to train a generative recurrent neural network, which can model RL environments through compressed spatiotemporal representations. It feeds extracted features into simple and compact policies, achieving impressive results in several environments.
Value propagation (VProp) \cite{Nardelli2018ValuePN} bases on value iteration, and is an efficient differentiable planning module.
It can successfully be trained to learn to plan using reinforcement learning.
As a general framework of AlphaZero,  MuZero\cite{deepmind2019muzreo} combines MCTS with a learned model, and predicts the reward, the action-selection policy, and the value function to make planning. It extends model-based RL to a range of logically complex and visually complex domains, and achieves superhuman performance. %Model-Based MetaPolicy-Optimization (MB-MPO) \cite{clavera2018model-based} is an model-based approach that foregoes the strong reliance on accurate learned dynamics models. Using an ensemble of learned dynamic models, MB-MPO meta-learns a policy that can quickly adapt to any model in the ensemble with one policy gradient step.

A general review of various DRL methods from 2017 to 2019 is presented in Table I.
\begin{table*}[!t]
\renewcommand{\arraystretch}{1.3}
\caption{A general review of recent DRL methods from 2017 to 2018.}
\label{table_example}
\centering
\begin{tabular}{c  c  c  c c }
\hline
\hline
DRL Algorithms & Main Techniques & Networks & Category \\
\hline
DQN \cite{Mnih2015Human} & experience replay, target Q-network & CNN  &  value-based, off-policy \\
Double DQN \cite{Hasselt2016DeepRL} & double Q-learning  & CNN  &  value-based, off-policy \\
Dueling DQN \cite{Wang2016DuelingNA} & dueling neural network architecture & CNN  &  value-based, off-policy  \\
Prioritized DQN \cite{Schaul2015PrioritizedER} & prioritized experience replay  & CNN  &  value-based, off-policy  \\
Bootstrapped DQN \cite{Osband2016DeepEV} & combine deep exploration with DNNs & CNN  &  value-based, off-policy  \\
Gorila \cite{Nair2015MassivelyPM} & massively distributed architecture  & CNN  &  value-based, off-policy  \\
LS-DQN \cite{Levine2017ShallowUF} & combine least-squares updates in DRL & CNN  &  value-based, off-policy  \\
Averaged-DQN \cite{Anschel2017AveragedDQNVR} & averaging learned Q-values estimates & CNN & value-based, off-policy  \\
DQfD \cite{hester2017deep} & learn from the demonstration data & CNN &  value-based, off-policy  \\
DQN with Pop-Art \cite{Hasselt2016LearningVA} & adaptive normalization with Pop-Art & CNN & value-based, off-policy  \\
Soft DQN \cite{Schulman2017EquivalenceBP}   & KL penalty and entropy bonus & CNN &  value-based, off-policy  \\
DQV \cite{sabatelli2018deep} & training a Quality-value network & CNN  & value-based, off-policy  \\
Rainbow  \cite{Hessel2018RainbowCI} & integrate six extensions to DQN & CNN & value-based, off-policy \\
RUDDER \cite{ArjonaMedina2018RUDDERRD} & return decomposition  & CNN-LSTM & value-based, off-policy  \\
Ape-X DQfD \cite{Pohlen2018ObserveAL} & transformed Bellman operator, temporal consistency loss & CNN & value-based, off-policy \\
C51 \cite{Bellemare2017A} & distributional Bellman optimality & CNN & value-based, off-policy \\
QR-DQN \cite{Dabney2018DistributionalRL} & distributional RL with Quantile regression & CNN & value-based, off-policy \\
IQN \cite{Dabney2018ImplicitQN} & an implicit representation of the return distribution & CNN  & value-based, off-policy \\
A3C \cite{Mnih2016AsynchronousMF} & asynchronous gradient descent & CNN-LSTM & policy gradient, on-policy  \\
GA3C \cite{Babaeizadeh2016ReinforcementLT} & hybrid CPU/GPU version & CNN-LSTM & policy gradient, on-policy \\
PPO \cite{Schulman2017ProximalPO} & clipped surrogate objective, adaptive KL penalty coefficient & CNN-LSTM & policy gradient, on-policy \\
ACER \cite{Wang2016SampleEA} & experience replay, truncated importance sampling  & CNN-LSTM & policy gradient, off-policy \\
ACKTR \cite{Wu2017ScalableTM} & K-FAC with trust region & CNN-LSTM & policy gradient, on-policy \\
Soft Actor-Critic \cite{haarnoja2018soft}   &   entropy regularization & CNN &  policy gradient, off-policy  \\
UNREAL \cite{Jaderberg2017ReinforcementLW} & unsupervised auxiliary tasks &  CNN-LSTM & policy gradient, on-policy  \\
Reactor \cite{Gruslys2017TheRA} & Retrace($\lambda$), $\beta$-leave-one-out policy gradient estimate & CNN-LSTM & policy gradient, off-policy \\
PAAC \cite{Clemente2017EfficientPM} & parallel framework for A3C & CNN & policy gradient, on-policy  \\
DDPG \cite{Lillicrap2015ContinuousCW} & DQN with deterministic policy gradient  & CNN-LSTM & policy gradient, off-policy \\
TRPO \cite{Schulman2015TrustRP} & incorporate a KL divergence constraint   & CNN-LSTM & policy gradient , on-policy \\
D4PG \cite{BarthMaron2018DistributedDD} & distributed distributional DDPG & CNN &  policy gradient , on-policy  \\
PGQ  \cite{ODonoghue2016PGQCP} & combine policy gradient and Q-learning & CNN &  policy gradient, off-policy   \\
IMPALA \cite{Espeholt2018IMPALASD} & importance-weighted actor learner architecture & CNN-LSTM & policy gradient, on-policy \\
FiGAR-A3C \cite{Sharma2017LearningTR} & fine grained action repetition & CNN-LSTM & policy gradient, on-policy  \\
%NoisyNet-DQN/A3C \cite{Fortunato2017NoisyNF} & parametric noise added in weights & CNN-LSTM & value-based/policy gradient \\
%FAR-AQL/A3C \cite{Sharma2017LearningTF} & factored action space representations & CNN-LSTM & value-based/policy gradient , off-policy \\
%Ape-X DQN/DPG \cite{Horgan2018DistributedPE} & distributed prioritized experience replay & CNN & value-based/policy gradient , off-policy  \\
TreeQN/ATreeC \cite{Farquhar2017TreeQNAA} &  on-line planning, tree-structured model & CNN &  model-based, on-policy  \\
STRAW \cite{Vezhnevets2016StrategicAW} & macro-actions, planning strategies & CNN & model-based, on-policy  \\
World model \cite{ha2018recurrent} & mixture density network, variational autoencoder  & CNN-LSTM & model-based, on-policy  \\
MuZero \cite{deepmind2019muzreo} & representation function,  dynamics function,  and prediction function & CNN & model-based, off-policy  \\
\hline
\hline
\end{tabular}
\end{table*}

\begin{figure*}[!t]
\centering
\includegraphics[width=6 in]{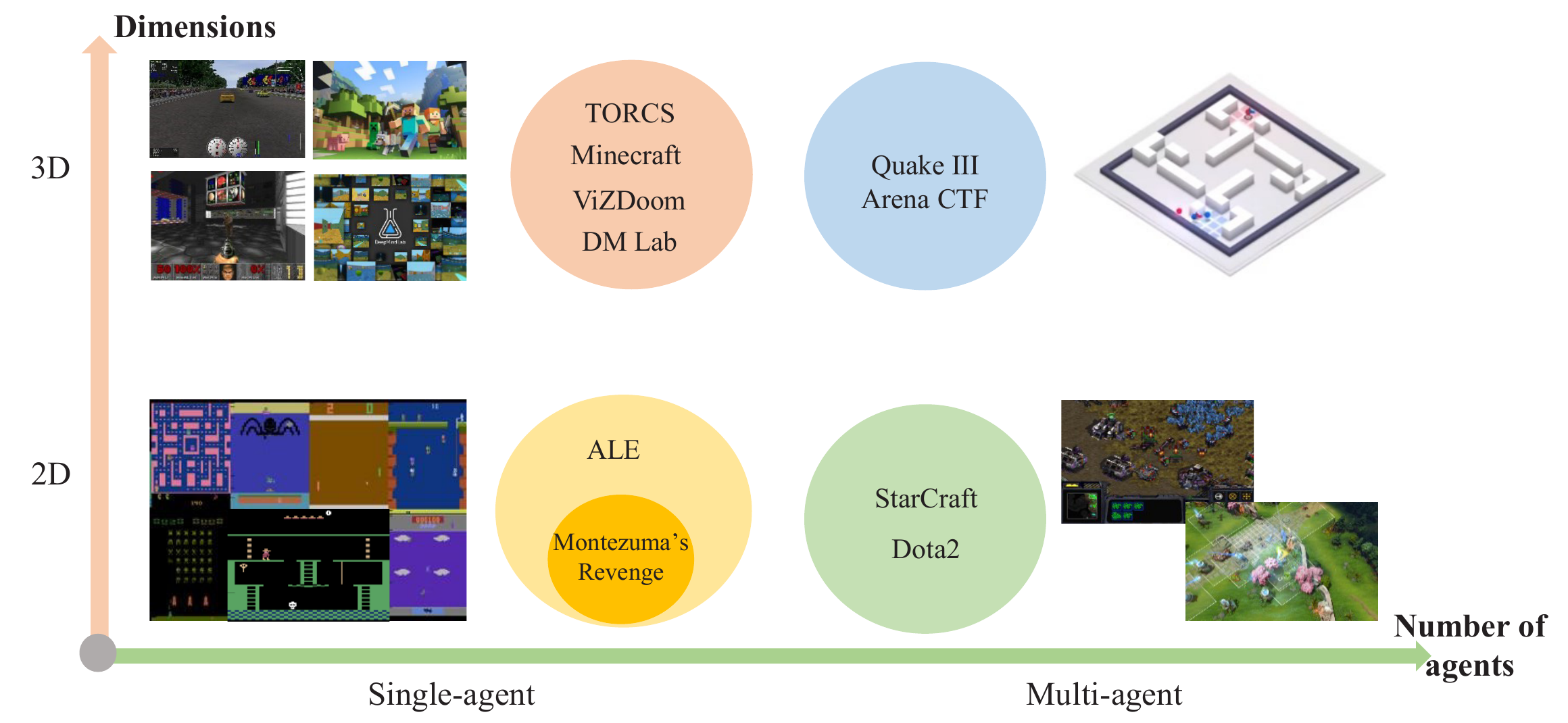}
\caption{The diagram of various video games AI, from 2D to 3D, and from single-agent to multi-agent.}
\label{fig_sim}
\end{figure*}

\section{DRL in video games}
Playing video games like human experts is challenging for computers.
With the development of DRL, agents are able to play various games end-to-end.
%In this field, game AI has made impressive progress on chess and Go, and now has moved to more complex video games, e.g. StarCraft and Dota2.
Here we focus on game research platforms and competitions, and impressive progress in various video games, from 2D to 3D, and from single-agent to multi-agent, as shown in Fig. 3.

\subsection{Game research platforms}

\begin{table}[!t]
\renewcommand{\arraystretch}{1.3}
\caption{A list of game AI competitions suitable for DRL research.}
\label{table_example}
\centering
\begin{tabular}{c c c c c c}
\hline
\hline
Competition Name & Time \\
\hline
ViZDoom AI competition & 2016, 2017, 2018 \\
StarCraft AI competitions (AIIDE, CIG, SSCAIT) & 2010 --- 2019 \\
microRTS competition & 2017, 2018, 2019 \\
The GVGAI competition -- learning track & 2017, 2018 \\
Microsoft Malmo collaborative AI challenge & 2017 \\
The multi-agent RL in Malmo competition & 2018 \\
The OpenAI Retro contest & 2018 \\
NeurIPS Pommerman competition & 2018 \\
Unity Obstacle Tower Challenge & 2019 \\
NeurIPS MineRL competition & 2019 \\
\hline
\hline
\end{tabular}
\end{table}

Platforms and competitions make great contributions  to the development of game AI, and help to evaluate agents' intelligence, as presented in Table II.
Most platforms can be described by two major categories: General Platforms and Specific Platforms.

\textbf{General Platforms:} Arcade Learning Environment (ALE) \cite{bellemare2013ale} is the pioneer evaluation platform for DRL algorithms, which provides an interface to plenty of Atari 2600 games. ALE presents both game images and signals, such as player scores, which makes it a suitable testbed. To promote the progress of DRL research, OpenAI integrates a collection of reinforcement learning tasks into a platform called Gym \cite{brockman2016openai}, which mainly contains Algorithmic, Atari, Classical Control, Board games, 2D and 3D robots. After that, OpenAI Universe \cite{openai2018universe} is a platform for measuring and training agents' general intelligence across a large supply of games. Gym Retro \cite{openai2018Retro} is a wrapper for video game emulator with a unified interface as Gym, and makes Gym easy to be extended with a large collection of video games, not only Atari but also NEC, Nintendo, and Sega, for RL research. The OpenAI Retro contest aims at exploring the development of DRL that can generalize from previous experience. OpenAI bases on the \textit{Sonic the Hedgehog$^{TM}$} video game, and presents a new DRL benchmark \cite{Nichol2018GottaLF}. This benchmark can help to measure the performance of few-shot learning and transfer learning in reinforcement learning. General Video Game Playing \cite{PrezLibana2018GeneralVG} is intended to design an agent to play multiple video games without human intervention. The General Video Game AI (GVGAI) \cite{Torrado2018DeepRL} competition is proposed to provide a easy-to-use and open-source platform for evaluating AI methods, including DRL. DeepMind Lab \cite{beattie2016deepmindlab} is a first-person perspective learning environment, and provides multiple complicated tasks in  partially observed, large-scale, and visually diverse worlds.
Unity ML-Agents Toolkit\cite{Juliani2018Unity} is a new toolkit for creating and interacting with simulation environments.
This platform has sensory, physical, cognitive, and social complexity, and enables fast and distributed simulation, and flexible control.

\textbf{Specific Platforms:}  Malmo\cite{microsoft2017Malmo} is a research platform for AI experiments, which is built on top of Minecraft. It is a first-person 3D environment, and can be used for multi-agent research in Microsoft Malmo collaborative AI challenge 2017 and the multi-agent RL in MalmO competition 2018. TORCS \cite{wymann2000torcs} is a racing car simulator which has both low-level and visual features for the self-driving car with DRL. ViZDoom \cite{Kempka2017ViZDoom} is a first-person shooter game platform, and encourages DRL agent to utilize the visual information to perform navigation and shooting tasks in a semi-realistic 3D world. ViZDoom AI competition has attracted plenty of researchers to develop their DRL-based Doom agents since 2016. As far as we know, real-time strategy (RTS) games are very challenging for reinforcement learning method. Facebook proposes TorchCraft for StarCraft I \cite{Synnaeve2016TorchCraftAL}, and DeepMind releases StarCraft II learning environment \cite{Vinyals2017StarCraftIA}. They expect researchers to propose powerful DRL agents to achieve high-level performance in RTS games and  annual StarCraft AI competitions.
CoinRun \cite{Cobbe2018Quantifying} provides a metric for an agent's ability to transfer its experience to novel situations. This new training environment strikes a desirable balance in complexity: the environment is much simpler than traditional platform games, but it still poses a worthy generalization challenge for DRL algorithms.
Google Research Football is a new environment based on open-source game Gameplay Football for DRL research.

\subsection{Atari games}
ALE is an evaluation platform that aims at building agents with general intelligence across hundreds of Atari 2600 games. As the most popular testbed for DRL research, a large number of DRL methods have achieved outstanding performance consecutively. Machado et al. \cite{Machado2018RevisitingTA} takes a review at the ALE in DRL research community, proposes diverse evaluation methodologies and some key concerns.
In this section, we will introduce the main achievements in the ALE domain, including the extremely difficult Montezuma's Revenge.

As the milestone in this domain, DQN is able to surpass the performances of previous algorithms, and achieves human-level performance across 49 games \cite{Mnih2015Human}.
Averaged-DQN examines the source of value function estimation errors, and demonstrates significantly improved stability and performance on the ALE benchmark \cite{Anschel2017AveragedDQNVR}.
UNREAL significantly outperforms the previous best performance on Atari, averaging 880\% expert human performance \cite{Jaderberg2017ReinforcementLW}.
PAAC achieves sufficiently good performance on ALE after a few hours of training \cite{Clemente2017EfficientPM}.
%Retrace($\lambda$) shows some benefits on a standard suite of Atari 2600 games, and Reactor achieves a new state-of-the-art performance on 57 games.
DQfD has better initial performance than DQN on most Atari games, and receives more average rewards than DQN on 27 of 42. In addition, DQfD learns faster than DQN even when given poor demonstration data \cite{hester2017deep}.
Noisy DQN replaces the conventional exploration heuristics with NoisyNet, and yields substantially higher scores in ALE domain.
As a distributional DRL method, C51 obtains a new series of impressive results, and demonstrates the importance of the value distribution in approximated RL \cite{Bellemare2017A}.
Rainbow provides improvements in terms of sample efficiency and final performance. The authors also show the contribution of each component to overall performance \cite{Hessel2018RainbowCI}.
QR-DQN algorithm significantly outperforms recent improvements on DQN, including the related C51 \cite{Bellemare2017A}.
IQN shows substantial gains on the Atari benchmark over QR-DQN, and even halves the distance between QR-DQN and Rainbow \cite{Dabney2018ImplicitQN}.
%RUDDER outperforms many DRL methods on the delayed reward Atari game Venture in only a fraction of the learning time. RUDDER considerably improves the state-of-the-art on the delayed reward Atari game Bowling in much less learning time.
Ape-X DQN substantially improves the performance on the ALE, achieving better final score in less wall-clock training time \cite{Horgan2018DistributedPE}.
When tested on a set of 42 Atari games, the Ape-X DQfD algorithm exceeds the performance of an average human on 40 games using a common set of hyperparameters.
Mean and median scores across multiple Atari games of typical DRL methods that achieve state-of-the-art performance consecutively are presented in Table III.

\begin{table}[!t]
\renewcommand{\arraystretch}{1.3}
\caption{Mean and median scores across 57 Atari games of typical DRL methods, measured as percentages of human baseline. }
\label{table_example}
\centering
\begin{tabular}{c c c c c c}
\hline
\hline
Methods & Mean & Median & year\\
\hline
DQN \cite{Mnih2015Human} &  228\% & 79\% & 2015\\
% Prioritized DQN \cite{Schaul2015PrioritizedER} & 434\% & 124\% & 2015\\
% Double DQN \cite{Hasselt2016DeepRL} & 307\% & 118\% & 2016\\
% Dueling DQN \cite{Wang2016DuelingNA} & 373\% & 151\% & 2016\\
% Prioritized dueling DQN \cite{Wang2016DuelingNA} & 592\% & 172\% & 2016\\
C51 \cite{Bellemare2017A} & 701\% & 178\% & 2017\\
UNREAL \cite{Jaderberg2017ReinforcementLW} & 880\% & 250\% & 2017\\
QR-DQN \cite{Bellemare2017A} & 915\% & 211\% & 2017\\
IQN \cite{Dabney2018ImplicitQN} & 1019\% & 218\% & 2018\\
Rainbow \cite{Hessel2018RainbowCI} & 1189\% & 230\% & 2018\\
% DQfD & 364\% & 113\% \\
% ACER & 570\% 144\% & \\
% Reactor & 555\% & 140\% \\
Ape-X DQN \cite{Horgan2018DistributedPE} & 1695\% & 434\% & 2018\\
Ape-X DQfD \makebox{$^{*}$}\cite{Pohlen2018ObserveAL} & 2346\% & 702\% & 2018\\
\hline
\hline
\end{tabular} \\
\ \\
\small Note: \makebox{$^{*}$} means this method is measured across 42 Atari games.
\end{table}

Montezuma's Revenge is one of the most difficult Atari video games. It is a goal-directed behavior learning environment with long horizons and sparse reward feedback signals. Players must navigate through a number of different rooms, avoid obstacles and traps, climb ladders up and down, and then pick up the key to open new rooms. It requires a long sequence of actions before reaching the goal and receiving a reward, and is difficult to explore an optimal policy to tackle tasks.
Efficient exploration is considered as a crucial factor to learn in a delayed feedback environment.
%Bellemare et al. \cite{bellemare2016unifying} introduce count-based exploration and intrinsic motivation with specific double DQN to train agents, and reaches a best performance of 6600 points.
Then, Ostrovski et al. \cite{ostrovski2017count} provide an improved version of count-based exploration with PixelCNN as a supplement for pseudo-count, also reveals the importance of Monte Carlo return for effective exploration.
%Hierarchical reinforcement learning method is a direct way to reduce the exploration space. Kulkarni et al. \cite{kulkarni2016hierarchical} present a hierarchical-DQN framework which is constituted by top-level module and low-level module, and allows efficient exploration in a delayed and sparse reward environment.
In addition to improve the exploration efficiency, learning from human data is also a proper method to reach better performance in this problem.
Le et al. \cite{le2018hierarchical} leverage imitation learning from expert interaction and hierarchical reinforcement learning at different levels. This method learns obviously faster than original hierarchical reinforcement learning, and also significantly more efficiently than conventional imitation learning.
%Human checkpoint replay \cite{hosu2016playing} is used as starting points for learning process, and increases the feasibility to find successful control policies earlier.
Other than gameplay, the demonstration is also a valuable kind of sample for agent to learn. DQfD utilizes a small set of demonstration data to speed up the learning process \cite{hester2017deep}. It combines prioritized replay mechanism with temporal difference updates and supervised classification, and finally achieves a better and impressive result. Further, Aytar et al. \cite{aytar2018playing} only use YouTube video as a demonstration sample and invests a transformed Bellman operator for learning from human demonstrations. Interestingly, these two works both claim being the first to solve the entire first level of Montezuma's Revenge.
Go-explore \cite{uber2018Goexplore} makes further progress, and achieves scores over 400,000 on average. Go-Explore separates learning into exploration and robustification. It reliably solves the whole game, and generalizes well.

\subsection{First-person perspective games}
Different from Atari games, agents in first-person perspective video games can only receive observations from their own perspectives, resulting from imperfect information inputs. In RL domain, this is a POMDP problem which requires efficient exploration and memory.

\subsubsection{\textbf{ViZDoom}}
First-person shooter (FPS)  games play an important role in game AI research.
Doom is a classical FPS game, and ViZDoom is presented as a novel testbed for DRL \cite{Kempka2017ViZDoom}.
Agents learn from visual inputs, and interact with the ViZDoom environment in a first-person perspective.
Wu et al. \cite{Wu2017Training} propose a method that combines A3C and curriculum learning.
The agent learns to navigate and attack via playing against built-in agents progressively. %, and wins the first place for track 1 in ViZDoom AI competition 2016.
%Dosovitskiy et al. \cite{dosovitskiy2016learning} propose direct future prediction (DFP), which utilizes a high-dimensional sensory stream and a low-dimensional measurement stream as inputs to provide a rich supervisory signal, and trains a sensorimotor control model by interacting with environment.
% DFP successfully generalizes across environments and goals, and wins the first place for track 2 in ViZDoom AI competition 2016.
%In \cite{Amiranashvili2018TDON}, the authors re-examine the role of TD in modern DRL, and find that finite horizon MC methods are not inferior to TD, even in sparse or delayed reward tasks, making MC a viable alternative to TD. They discuss the role of perceptual complexity in reconciling these findings with classic empirical results.
Parisotto et al. \cite{Parisotto2017NeuralMS} develop Neural Map, which is a memory system with an adaptable write operator. Neural Map uses a spatially structured 2D memory image to store the environment's information. This method surpasses other DRL memories on several challenging ViZDoom maze tasks and shows a capable generalization ability.
Shao et al. \cite{shao2018battles} show that ACKTR can successfully teach agents to battle in ViZDoom environment, and significantly outperform A2C agents by a significant margin.

\subsubsection{\textbf{TORCS}}
TORCS is a racing game where actions are acceleration, braking and steering. %It has previously been used as a testbed in RL.
This game has more realistic graphics than Atari games, but also requires agents to learn the dynamic of the car.
%DDPG is able to learn reasonable policies to complete a circuit around the track in TORCS on both low-dimensional and visual inputs.
FIGAR-DDPG can successfully complete the race task and finish 20 laps of the circuit, with a 10$\times$ total reward against that obtained by DDPG, and much smoother policies \cite{Sharma2017LearningTR}.
Normalized Actor-Critic (NAC) normalizes the Q-function effectively, and learns an initial policy network from demonstration and refine the policy in a real environment \cite{Gao2018ReinforcementLF}. NAC is robust to suboptimal demonstration data, learns robustly and outperforms existing baselines when evaluated on TORCS.
Mazumder et al. \cite{Mazumder2018Action} incorporate state-action permissibility (SAP) and DDPG, and applies it to tackle the lane keeping problem in TORCS. The proposed method can speedup DRL training remarkably for this task.
In \cite{li2018cim}, a two-stage approach is proposed for the vision-based vehicle lateral control problem which includes an multi-task learning perception stage and an RL control stage. By exerting the correlation between multiple learning task, the perception module can robustly extract track features. Additionally, the RL agent learns by maximizing the geometry-based reward and performs better than the LQR and MPC controllers.
Zhu et al. \cite{zhu2018torcs} use DRL to train a CNN to perceive driving data from images of first-person view, and learns a controller to get driving commands, showing a promising performance.

\subsubsection{\textbf{Minecraft}}

Minecraft is a sandbox construction game, where players can build creative creations, structures, and artwork across various game modes.
Recently, it becomes a popular platform for game AI research, with 3D infinitely varied data.
Project Malmo is an experimentation platform \cite{Johnson2016TheMP} that builts on the Minecraft for AI research. It supports a large number of scenarios, including navigation, problem solving tasks, and survival to collaboration.
Xiong et al. \cite{Xiong2018HogRiderCA} propose a novel Q-learning approach with state-action abstraction and warm start using human reasoning to learn effective policies in the Microsoft Malmo collaborative AI challenge.
%Oh et al. \cite{Oh2016ControlOM} introduce a set of visual navigation tasks in Minecraft, and proposes a memory-based DRL architecture, termed as feedback recurrent memory Q-network (FRMQN). FRMQN can take control of memory, active perception and action. This architecture has a better generalization to unseen environments than existing DRL architectures.
The ability to transfer knowledge from source task to target task in Minecraft is one of the major challenges. Tessler et al. \cite{Tessler2017ADH} provides a DRL agent which can transfer knowledge by learning reusable skills, and then incorporated into hierarchical DRL network (H-DRLN). H-DRLN exhibits superior performance and low learning sample complexity compared to regular DQN in Minecraft, and the potential to transfer knowledge between related Minecraft tasks without any additional learning.
To solve the partial or non-Markovian observations problems,
Jin et al. \cite{Jin2017RegretMF} propose a new DRL algorithm based on counterfactual regret minimization that iteratively updates an approximation to a cumulative clipped advantage function.
On the challenging Minecraft first-person navigation benchmarks, this algorithm can substantially outperform strong baseline methods.

\subsubsection{\textbf{DeepMind lab}}
%\begin{table}[!t]
%\renewcommand{\arraystretch}{1.3}
%\caption{Average scores across DM-30 of typical model-free DRL methods.}
%\label{table_example}
%\centering
%\begin{tabular}{c c c c c c}
%\hline
%\hline
%Methods & Mean & Median & \\
%\hline
%Kickstarting &  228\% & 79\% \\
%
%\hline
%\hline
%\end{tabular}
%\end{table}
DeepMind lab is a 3D first-person game platform extended from OpenArena, which is based on Quake3.
%It is comparable to other first-person 3D game platforms for AI research.
Comparable to other first-person game platforms, DeepMind lab has considerably richer visuals and more realistic physics, making it a significantly complex platform.
On a challenging suite of DeepMind lab tasks, the UNREAL agent leads to a mean speedup in learning of 10$\times$ over A3C and averaging 87\% expert human performance.
As learning agents become more powerful, continual learning has made quick progress recently. To test continual learning capabilities, Mankowitz et al. \cite{Mankowitz2018UnicornCL} consider an implicit sequence of tasks with sparse rewards in DeepMind lab.
The novel agent architecture called Unicorn, demonstrates strong continual learning and outperforms several baseline agents on the proposed domain.
Schmitt et al. \cite{Schmitt2018KickstartingDR} present a method which uses teacher agents to kickstart the training of a new student agent. On a multi-task and challenging DMLab-30 suite, kickstarted training improves new agents' sample efficiency to a great extend, and surpasses the final performance by 42\%.
Jaderberg et al. \cite{Jaderberg2018HumanlevelPI} focus on Quake III Arena Capture the Flag, which is a popular 3D first-person multiplayer video game, and demonstrates that DRL agents can achieve human-level performance with only pixels and game points as input.
The agent uses population based training to optimize the policy. This method trains a large number of agents concurrently from thousands of parallel matches, where agents plays cooperatively in teams and against each other on randomly generated environments.
In an evaluation, the trained agents exceed the winrate of self-play baseline and high-level human players both as teammates and opponents, and are proved far stronger than existing DRL agents.

\subsection{Real-time strategy games}
Real-time strategy games are very popular among players, and have become popular platforms for AI research.
%Due to its popularity, RTS has attracted intensive interest from AI researchers.

\subsubsection{\textbf{StarCraft}}
In StarCraft, players need to perform actions according to real-time game states, and defeat the enemies.
Generally speaking, designing an AI bot have many challenges, including multi-agent collaboration, spatial and temporal reasoning, adversarial planning, and opponent modeling.
Currently, most bots are based on human experiences and replays, with limited flexibility and intelligence.
%DRL is good at environmental perception and sequential decisions.
DRL is proved to be a promising direction for StarCraft AI, especially in micromanagement, build order, mini-games and full-games \cite{tang2018starcraft}.

Recently, micromanagement is widely studied as the first step to solve StarCraft AI.
%RL agents in micromanagement have to tackle a lot of challenges, such as durative actions, delayed rewards, and large state and action spaces.
Usunier et al. \cite{Usunier2017Episodic} introduce the greedy MDP with episodic zero-order optimization (GMEZO) algorithm to tackle micromanagement scenarios, which performs better than DQN and policy gradient.
BiCNet \cite{Peng2017Multiagent} is a multi-agent deep reinforcement learning method to play StarCraft combat games.
It bases on actor-critic reinforcement learning, and uses bi-directional neural networks to learn collaboration.
BiCNet successfully learns some cooperative strategies, and is adaptable to various tasks, showing better performances than GMEZO.
In aforementioned works, researchers mainly develops centralized methods to play micromanagement.
Foerster et al. \cite{Foerster2017Stabilising} focus on decentralized control for micromanagement, and propose a multi-agent actor-critic method.
To stabilize experience replay and solve nonstationarity, they use fingerprints and importance sampling, which can improve the final performance.
Shao et al. \cite{shao2018starcraft} follow decentralized micromanagement task, and propose parameter sharing multi-agent gradient descent SARSA($\lambda$) (PS-MAGDS) method. To resue the knowledge between various micromanagement scenarios, they also combine curriculum transfer learning to this method.
This improves the sample efficiency, and outperforms GMEZO and BiCNet in large-scale scenarios.
%In decentralized tasks, COMA significantly improves average performance over other multi-agent actor-critic methods.
Kong et al. \cite{Kong2017RevisitingTM} bases on master-slave architecture, and proposes master-slave multi-agent reinforcement learning (MS-MARL).
MS-MARL includes composed action representation, independent reasoning, and learnable communication. This method has better performance than other methods in micromanagement tasks.
Rashid et al. \cite{Rashid2018QMIXMV} focus on several challenging StarCraft II micromanagement tasks, and use centralized training and decentralized execution to learn cooperative behaviors.
%QMIX estimates joint action-values as a complex non-linear combination of per-agent values. These values condition only on local observations through a neural network. The authors evaluate QMIX
This eventually outperforms state-of-the-art multi-agent deep reinforcement learning methods.

Researchers also use DRL methods to optimize the build order in StarCraft. Tang et al. \cite{Tang2018Reinforcement} put forward neural network fitted Q-learning (NNFQ) and convolutional neural network fitted Q-learning (CNNFQ) to build units in simple StarCraft maps. These models are able to find effective production sequences, and eventually defeat enemies.
In \cite{Vinyals2017StarCraftIA}, researchers present baseline results of several main DRL agents in the StarCraft II domain.
The fully convolutional advantage actor-critic (FullyConv-A2C) agents achieve a beginner-level in StarCraft II mini-games.
%However, these agents are unable to make significant progress on the main game.
%\cite{Sukhbaatar2017IntrinsicMA} proposes asymmetric self-play to build Marine units, which greatly speeds up learning, and surpasses the count-based approach.
Zambaldi et al. \cite{Zambaldi2018RelationalDR} introduce the relational DRL to StarCraft, which iteratively reasons about the relations between entities with self-attention, and uses it to guide a model-free RL policy.
This method improves sample efficiency, generalization ability, and interpretability of conventional DRL approaches.
Relational DRL agent achieves impressive performance on SC2LE mini-games.
Sun et al. \cite{Sun2018TStarBot} develop the DRL based agent TStarBot, which uses flat action structure. This agent defeats the built-in AI agents from level 1 to level 10 in a full game firstly.
Lee et al. \cite{Dennis2018Modular}  focus on StarCraft II AI, and present a novel modular architecture, which splits responsibilities between multiple modules. Each module controls one aspect of the game, and two modules are trained with self-play DRL methods. This method defeats the built-in bot in "Harder" level.
Pang et al. \cite{Pang2018On} investigate a two-level hierarchical RL approach for StarCraft II. The macro-action is automatically extracted from expert's data, and the other is a flexible and scaleable hierarchical architecture. More recently, DeepMind proposes AlphaStar, and defeats professional players for the first time.

%\begin{figure*}[!t]
%\centering
%\includegraphics[width=6 in]{points.pdf}
%\caption{Key points and research directions of DRL in video games.}
%\label{fig_sim}
%\end{figure*}

\subsubsection{ \textbf{MOBA and Dota2}}

MOBA (Multiplayer Online Battle Arena) is originated from RTS games, which has two teams, and each team consists of five players. To beat the opponent, five players in a team must cooperate together, kill enemies, upgrade heros, and eventually destroy the opponent base.
Since MOBA research is still in a primary stage, there are fewer works than conventional RTS games. Most works on MOBA concentrate on dataset analysis and case study. However, due to a series of breakthroughs that DRL achieves in game AI, researchers start to pay more attention to MOBA recently. King of Glory (a simplified mobile version of Dota) is the most popular mobile-end MOBA game in China. Jiang et al. \cite{jiang2018feedback} apply Monte-Carlo Tree Search and deep neural networks to this game. The experimental results indicate that MCTS-based DRL method is efficient and can be used in 1v1 MOBA scenario. Most impressive works on MOBA are proposed by OpenAI. Their results prove that DRL method with self-play can not only be successful in a 1v1 and 2v2 Dota2 scenarios \cite{openai2017Dota1}, but also in 5v5 \cite{openai2018DotaFive}\cite{openai2019dota2}. The model architecture is simple, using a LSTM layer as the core component of neural network. Under the support of massively distributed cloud computing and PPO optimization algorithm, OpenAI Five can master the critical abilities of team fighting, searching forest, focusing, chasing, and diversion for team victory, and defeat human champion OG with 2:0. Their works truly open a new door to MOBA research with DRL method.

% \section{Key points and research directions}
\section{Challenges in Games with DRL}

Since DRL has achieved large progress in some video games, it is considered as one of most promising ways to realize the artificial general intelligence. However, there are still some challenges should be conquered towards goal.  In this secition, we discuss some crucial challenges for DRL in video games, such as tradeoff between exploration and exploitation, low sample efficiency,  dilemma in generalization and overfiting,  multi-agent learning, incomplete information and delayed sparse rewards. Though there are some proposed approaches have been tried to solve these problems, as presented in Fig. 4, there are still some limitations should be broken.

%
%At present, there are still some major problems when applying DRL methods to this field, especially in 3D imperfect information multi-agent video games.
%In this section, we discuss these key points from different aspects, including exploration-exploitation, sample efficiency, generalization and transfer ability, multi-agent learning, imperfect information, and delayed spare rewards.
%Based on these key points, there are also some research directions proposed to solve these problems, as presented in Fig. 4.

\subsection{Exploration-exploitation}

Exploration can help to obtain more diversity samples, while exploitation is the way to learn the high reward policy with valuable samples.  The trade-off between exploration and exploitation remains a major challenge for RL. Common methods for exploration require a large amount of data, and can not tackle temporally-extended exploration. Most model-free RL algorithms are not computationally tractable in complicated environments.

%\subsubsection{\textbf{Noise and randomization}}
Parametric noise can help exploration to a large extend in the training process \cite{Fortunato2017NoisyNF} \cite{Plappert2017ParameterSN}.
Besides, randomized value functions become an effective approach for efficient exploration. Combining exploration with deep neural networks can help to learn much faster, which greatly improves the learning speed and final performance in most games \cite{Osband2016DeepEV}.

%\subsubsection{\textbf{Count-based approach}}
A simple generalization of popular count-based approach can reach satisfactory performance on various high-dimensional DRL benchmarks \cite{Tang2017ExplorationAS}. This method maps states to hash codes, and counts their occurrences via a hash table. Then,  according to the classic count-based method, we can use these counts to compute a reward bonus. On many challenging tasks, these simple hash functions can achieve impressive performance. This exploration strategy provides a simple and powerful baseline to solve MDPs requiring considerable exploration.

\subsection{Sample efficiency}
DRL algorithms usually take millions of samples to achieve human-level performance. While humans can quickly master highly rewarding actions of an environment.
Most model-free DRL algorithms are data inefficient, especially for a environment with high dimension and large explore space. 
They have to interact with environment in a large time cost for seek out high reward experiences in a complex sample space, which limits their applicability to many scenarios.
In order to reduce the exploration dimension of environment  and ease the expenditure of time on interaction, 
some solutions can be used for improving data efficiency, such as hierarchy and demonstration.

%\subsubsection{\textbf{Hierarchy}}
%Hierarchical reinforcement learning (HRL) tackles this problem by using a set of temporally-extended actions, or options.
Hierarchical reinforcement learning (HRL) allows agents to decompose the task into several simple subtasks, which can speed up training and improve sample efficiency.
Temporal abstraction is key to scaling up learning, while creating such abstractions autonomously has remained challenging.
The option-critic architecture has the ability to learn the internal policies and the options' termination conditions, without any additional rewards or subgoals \cite{Bacon2017TheOA}.
FeUdal Networks (FuNs) include a Manager module and a Worker module \cite{Vezhnevets2017FeUdalNF}. The Manager sets abstract goals at high-level. The Worker receives these goals, and generates actions in the environment. FuN dramatically outperforms baseline agents on tasks that involve long-term credit assignment or memorization.
Representation learning methods can also be used to guide the option discovery process in HRL domain \cite{Machado2017EigenoptionDT}.

%\subsubsection{\textbf{Demonstration}}
Demonstration is a proper technique to improve sample efficiency.
Current approaches that learn from demonstration use supervised learning on expert data and use reinforcement learning to improve the performance. This method is difficult to jointly optimize divergent losses, and is very sensitive to noisy demonstrations.
Leveraging data from previous control of the system can greatly accelerate the learning process even with small amounts of demonstration data \cite{hester2017deep}.
Goals defined with human preferences can effectively solve complicated RL tasks without the reward function, while greatly reducing the cost of human oversight \cite{Christiano2017DeepRL}. %This method can successfully train complex novel behaviors with less time.

\subsection{Generalization and Transfer}

%One of the main challenges in reinforcement learning is generalization. 
The ability to transfer knowledge across  multiple environments is considered as a critical aspect of intelligent agents.
%In addition, learning to solve complex sequences of tasks while leveraging transfer remains a key obstacle to achieve human-level intelligence.
With the purpose of promoting the performance of generalization in multiple environments, multi-task learning and policy distillation have been focus on these situations.

%\subsubsection{\textbf{Multi-task learning}}
Multi-task learning with shared neural network parameters can solve the generalization problem, and efficiency can be improved through transfer across related tasks.
Hybrid reward architecture takes a decomposed reward function as input and learns a separate value function for each component \cite{Seijen2017HybridRA}. The whole value function is much smoother, which can be easily approximated with a low-dimensional representation, and learns more effectively.
%Actor-Mimic \cite{Parisotto2016ActorMimicDM} is a novel method of transfer learning that enables an agent learning in multiple tasks simultaneously, and then generalizing its knowledge to new domains. This method exploits the use of DRL and model compression techniques to learn how to act in a set of distinct tasks by using the guidance of several expert teachers. The representations learnt by the deep policy network are capable of generalizing to new tasks with no prior expert guidance, speeding up learning in novel environments.
IMPALA shows the effectiveness for multi-task reinforcement learning, using less data and exhibiting positive transfer between tasks \cite{Espeholt2018IMPALASD} .
PopArt-IMPALA combines PopArt's adaptive normalization with IMPALA, and allows a more efficient use of parallel data generation, showing impressive performance on multi-task domain \cite{Hessel2018Multi-task}.
%PopArt-IMPALA results in state-of-the-art performance on learning to play all 57 Atari games, which exceeds median human performance. This is the first time a single agent surpassed human-level performance on this multi-task domain.

%\subsubsection{\textbf{Avoiding catastrophic forgetting}}
%As for avoiding catastrophic forgetting, the progressive neural network represents a step forward in this field \cite{Rusu2016ProgressiveNN}.
%This approach is immune to forgetting and can leverage prior knowledge via lateral connections to previously learned features.
%For artificial general intelligence (AGI), it will be efficient if multiple users train the same giant neural network, permitting parameter reuse, without catastrophic forgetting. PathNet \cite{Fernando2017PathNetEC} uses agents embedded in the neural network whose task is to discover the valuable parts for new tasks. %Agents are pathways through the network which determines the subset of parameters that are used and updated by the algorithm.
%Positive transfer is demonstrated for a set of reinforcement learning tasks, suggesting PathNet has general applicability for neural network training.
%%Most DRL algorithms are data inefficient in complex and rich environments, limiting their applicability to many scenarios.
%%Multitask learning with shared neural network parameters can improve data efficiency, where efficiency may be improved through transfer across related tasks.

%\subsubsection{\textbf{Policy distillation}}
To successfully learn complex tasks with DRL, we usually need large task-specific networks and extensive training to achieve good performance.
% A novel method called policy distillation \cite{Rusu2016PolicyD} is presented that can be used to extract the policy of a RL agent and train a new network that performs at the expert level while being dramatically smaller and more efficient.
%Teh et al.  propose a new approach for joint training of multiple tasks, termed as
Distral shares a distilled policy which can learn common knowledge across multiple tasks \cite{Teh2017DistralRM}. Each worker is trained to solve individual task and to be close to the shared policy, while the shared policy is trained by distillation. This approach shows efficient transfer on complex tasks, with more robust and more stable performance.
Mix \& Match is a training framework that is designed to encourage effective and rapid learning in DRL agents \cite{Czarnecki2018MixMatchA}. It allows to automatically form a curriculum over agent, and progressively trains more complex agents from simpler agents.

\subsection{Multi-agent learning}
Multi-agent learning is very important in video games, such as StarCraft.
In a cooperative multi-agent setting, curse-of-dimensionality, communication, and credit assignment are major challenges.

%\subsubsection{\textbf{Centralized training with decentralized execution}} 
%Q-learning is challenged by an inherent non-stationarity of the environment, while policy gradient suffers from a variance that increases as the number of agents grows.
Team learning uses a single learner to learn joint solutions in multi-agent system, while concurrent learning uses multiple learners for each agent. 
Recently, the centralised training of decentralised policies is becoming a standard paradigm for multi-agent training.
Multi-agent DDPG considers other agents' action policy and can successfully learn complex multi-agent coordination behavior \cite{ryan2017multi}.
Counterfactual multi-agent policy gradients uses a centralized critic to estimate the action-value function and decentralized actors to optimize each agents' policies, with a counterfactual advantage function to address the multi-agent credit assignment problem \cite{Foerster2017Counterfactual} .
In addition, communication protocols is important to share information to solve multi-agent tasks. Reinforced Inter-Agent
Learning (RIAL) and Differentiable Inter-Agent Learning (DIAL) use deep reinforcement learning to learn end-to-end communication protocols in complex environments. Analogously, CommNet is able to learn continuous communication between multiple agents.
%\subsubsection{\textbf{Mean field}}
%%Communication helps the collaboration of multiple agents during training.
%%Backpropagation is able to learn communication in fully cooperative multi-agent environments \cite{sukhbaatar2016learning}.
%
%Multi-agent learning becomes intractable due to the curse of the dimensionality and the exponential growth of agent interactions.
%Mean Field Reinforcement Learning where
%The interactions within the population of agents are approximated by those between a single agent and the average effect from the overall population or neighboring agents.
%the interplay between the two entities is mutually reinforced: the learning of the individual agent's optimal policy depends on the dynamics of the population, while the dynamics of the population change according to the collective patterns of the individual policies.

\subsection{Imperfect information}
In partially observable and first-perspective games, DRL agents need to tackle imperfect information to learn a suitable policy.
Making decisions in these environments is challenging for DRL agents.

%\subsubsection{\textbf{Memory}}
A critical component of enabling effective learning in these environment is the use of memory.
DRL agents have used some simple memory architectures, such as several past frames or an LSTM layer. But these architectures are limited to only remember transitory information.
Model-free episode control learns difficult sequential decision-making tasks much faster, and achieves a higher overall reward \cite{Blundell2016ModelFreeEC}.
Differentiable neural computer uses a neural network  to read from and write to an external memory matrix \cite{Graves2016HybridCU}.
%When trained with reinforcement learning, a DNC can complete a moving blocks puzzle in which changing goals are specified by sequences of symbols.
This method can solve complex, structured tasks which can not access to neural networks without external read and write memory.
Neural episodic control inserts recent state representations paired with corresponding value functions into the appropriate neural dictionary, and learns significantly faster than other baseline agents \cite{Pritzel2017NeuralEC}.
%NEC agent uses a semi-tabular representation of the value function: a buffer of past experience containing slowly changing state representations and rapidly updated estimates of the value function.
%The core of NEC is a memory structure, where NEC inserts recent state representations paired with corresponding value functions into the appropriate neural dictionary.

\subsection{Delayed spare rewards}
The sparse and delayed reward is very common in many games, and is also one of the reasons that reduce sample efficiency in reinforcement learning.

%\subsubsection{\textbf{Curiosity}}
In many scenarios, researchers use curiosity as an intrinsic reward to encourage agents to explore environment and learn useful skills. Curiosity can be formulated as the error that the agent predicts its own actions' consequence in a visual space \cite{Pathak2017CuriosityDrivenEB}. This can scale to high-dimensional continuous state spaces. Moreover, it leaves out the aspects of environment that cannot affect agents.
Curiosity search for DRL encourages intra-life exploration by rewarding agents for visiting as many different states as possible within each episode \cite{Stanton2018DeepCS}.

\section{Conclusion and discussion}

Game AI with deep reinforcement learning  is a challenging and promising direction. Recent progress in this domain has promote the development of artificial intelligence research.
In this paper, we review the achievements of deep reinforcement learning in video games. Different DRL methods and their successful applications are introduced.
These DRL agents achieve human-level or super-human performances in various games, from 2D perfect information to 3D imperfect information, and from single-agent to multi-agent.
In addition to these achievements, there are still some major problems when applying DRL methods to this field, especially in 3D imperfect information multi-agent video game.
A high-level game AI requires to explore more efficient and robust DRL techniques, and needs novel frameworks to be implemented in complex environment.
These challenges have not been fully investigated and could be opened for further study in the future.
%The research theme of game AI is far more than the game itself. It can both bridge the gap between virtual environment and real world, and contribute to our daily life. DRL is proved to be a promising way to achieve this goal.

\section*{Acknowledgment}
The authors would like to thank Qichao Zhang, Dong Li and Weifan Li for the helpful comments and discussions about this work.

\ifCLASSOPTIONcaptionsoff
  \newpage
\fi

\bibliographystyle{IEEEtran}
\bibliography{refer-rl}

\end{document}